\pdfoutput=1

\documentclass[aps,pre,reprint,groupedaddress, %
  showpacs,amssymb,amsfonts,floatfix]{revtex4-1}

  \usepackage{graphicx}

  \graphicspath{{fig/}{./}}
  \DeclareGraphicsExtensions{.pdf,.jpg,.png}

  \usepackage[caption=false,font=footnotesize]{subfig}

  \usepackage{url}

  \usepackage[usenames,dvipsnames]{color}
  \usepackage[hyperfootnotes=false]{hyperref}
  \hypersetup{
     colorlinks=true,
     citecolor=Violet,
     linkcolor=Red,
     urlcolor=Blue}

  \usepackage{amssymb}
  \usepackage{amsmath}

  \usepackage{mathrsfs}

  \usepackage{bm}
  \newcommand{\vect}[1]{\bm{#1}}
  \newcommand{\vectb}[1]{\bm{#1}}

  \newcommand{\dd}{\; \mathrm{d}}

  \newcommand{\N}{\mathcal{N}} 

  \newcommand{\T}{^{\mathsf{T}}}

  \newcommand{\R}{\mathbb{R}}
  \newcommand{\E}{\mathrm{E}}

  \providecommand{\norm}[1]{\lVert#1\rVert}
  \providecommand{\op}[1]{\mathcal{#1}}

  \newcommand{\eg}{\textit{e.g.}}

  \newcommand{\cf}{\textit{cf.}}

  \newcommand{\ie}{\textit{i.e.}}

\begin{document}


  \title{Infinite-dimensional Bayesian filtering for detection \\
         of quasi-periodic phenomena in spatio-temporal data}

  \author{Arno Solin}
  \email[]{arno.solin@aalto.fi}
  \altaffiliation[Also at~]{Advanced Magnetic Imaging Centre (AMI),
                            Aalto University, Finland.}
  \affiliation{Department of Biomedical Engineering and Computational Science (BECS),
               Aalto University, P.O.~Box 12200, FI-00076~AALTO, Finland.}

  \author{Simo S\"arkk\"a}
  \email[]{simo.sarkka@aalto.fi}
  \altaffiliation[Also at~]{Advanced Magnetic Imaging Centre (AMI),
                            Aalto University, Finland.}
  \affiliation{Department of Biomedical Engineering and Computational Science (BECS),
               Aalto University, P.O.~Box 12200, FI-00076~AALTO, Finland.}

  \date{\today}



  \begin{abstract}
	This paper introduces a spatio-temporal resonator model and an inference
	method for detection and estimation of nearly periodic temporal
	phenomena in spatio-temporal data. The model is derived as a
	spatial extension of a stochastic harmonic resonator model, which
	can be formulated in terms of a stochastic differential equation
	(SDE). The spatial structure is included by introducing linear
	operators, which affect both the oscillations and damping, and by
	choosing the appropriate spatial covariance structure of the driving
	time-white noise process. With the choice of the linear
	operators as partial differential operators, the resonator
	model becomes a stochastic partial differential equation (SPDE),
	which is compatible with infinite-dimensional Kalman filtering.
	The resulting infinite-dimensional Kalman filtering problem
	allows for a computationally efficient solution as the
	computational cost scales linearly with measurements in the
	temporal dimension. This framework is applied to weather
	prediction and to physiological noise elimination in fMRI brain
	data.
  \end{abstract}

\maketitle

\section{Introduction}\label{sec:introduction}
Oscillations stem from repetitive variation, typically in time,
of some measure around a point or an equilibrium. This type of
phenomenon is commonly encountered in natural systems, as well as in
physical, biological, and chemical models \cite{Rotermund:1990,
Rzhanov:1993, Singh:2013}. This paper proposes a
computationally effective evolution-type stochastic partial
differential equation model and an inference method, which together provide a
novel and efficient means of detecting and modeling latent
oscillatory structures in space--time, such as physiological
noise in fMRI brain data \cite{Sarkka+Solin:2012,
Sarkka+Solin:2012b} or temperature variation in climate models.

The proposed model can be thought of as an extension of the
following simple stochastic harmonic resonator model (see,
\eg, \cite{Burrage+Lenane+Lythe:2008,Sarkka+Solin:2012}):
\begin{equation} \label{eq:oscillator}
  {\dd^2 f(t) \over \dd t^2} + \gamma {\dd f(t) \over \dd t}
    + \omega^2 f(t) = \xi(t),
\end{equation}
where $\xi(t)$ is temporally  white noise, $\gamma$ is the
damping coefficient, and the resonator frequency is defined by
the angular velocity $\omega$ (rad/s). Letting the
oscillation frequency change over time and including
harmonics allows the modeling of more complicated periodic and
quasi-periodic (almost periodic) properties
(\cf~\cite{Sarkka+Solin:2012, Hartikainen+Seppanen+Sarkka:2012}).

However, the oscillatory phenomena can also contain spatial
properties. This leads to a space--time model, where the process
can be described by a spatial field that is evolving in time.
The main contribution herein is to set up a model in which the temporal
behavior has oscillatory characteristics, such that the process
can be described by a spatio-temporal resonator model:
\begin{equation} \label{eq:SPDE-oscillator}
  {\partial^2 f(\vect{x},t) \over \partial t^2}
    + \op{A} {\partial f(\vect{x},t) \over \partial t}
    + \op{B} f(\vect{x},t) = \xi(\vect{x},t), \end{equation}
where $\op{A}$ and $\op{B}$ are linear operators modeling the
space--time interactions in the oscillator state. 
Additionally, it is shown
how infinite-dimensional Kalman filters and smoothers provide
computationally effective means of computing the Bayesian solution
for detecting and estimating the oscillations in noisy
measurement data.
This can be seen as a generalization of
diffusive coupling models \cite{Hale:1997} to stochastic oscillating
fields.

Previously, for discrete space and time, the spatio-temporal
interactions in such data have been modeled, for example, with
seasonal VARMA (vector autoregressive moving average)
(see, \eg, \cite{Box+Jenkins+Reinsel:2008,
Pindyck+Rubinfeld:1981}) models. However, incorporating the spatial
structure and predicting new measurements in space and time is
difficult, if not impossible, in these models. For
continuous-valued treatment, one can resort to neural networks
(see, \eg, \cite{Haykin:1999}), which make it possible to
account for the latent structure, but provide few tools for
assessing the model structure or interpreting the results.

It would also be possible to formulate the spatio-temporal model
and the related Bayesian estimation problem directly in terms of
Gaussian processes (GPs), for example by using periodic covariance
functions \cite{Rasmussen:2006}. Unfortunately, the direct
use of this approach leads to an intractable cubic computational
complexity $\mathcal{O}(T^3)$ in the number of time steps $T$.
To some extent, it is possible to reduce this problem by using
sparse approximations (see, \eg, \cite{Quinonero-Candela+Rasmussen:2005, Rasmussen:2006}), but this does not solve the problem fully.

The use of \emph{stochastic partial differential equation}
(SPDE) based models to form computationally efficient solutions to
Gaussian process regression problems (or equivalent Kriging
problems) has recently been discussed in
\cite{Lindgren+Rue+Linstrom:2011,Sarkka:AISTATS:2012}. In
particular, in \cite{Sarkka:AISTATS:2012} the authors propose a
method for converting a covariance function based spatio-temporal
Gaussian process regression model into an equivalent SPDE type
model. The advantage of this approach is that the Bayesian
inference problem of the resulting model can be solved using
infinite-dimensional Kalman filtering and smoothing with linear
computational complexity in time.

This work follows an approach similar to
\cite{Sarkka:AISTATS:2012}, except that the SPDE
model is formulated directly as a linear combination of spatio-temporal
oscillators, rather than first forming a covariance function-based
Gaussian process regression problem and then converting it into
an SPDE. Although some modeling freedom is lost in the present approach,
the advantage is that it always produces a model that can be
solved with a linear time complexity algorithm, and no additional
conversion procedures are required to achieve this.

The following sections introduce the spatio-temporal
resonator model and explain how to select the spatial
operators. The operators are chosen such that an
orthogonal basis for the model can be formed by using the eigenvalue
decomposition of the Laplace operator. The Hilbert space method
approach to solving the GP model using Kalman filtering is discussed 
in brief together with maximum likelihood parameter
estimation. As a proof-of-concept demonstration a
one-dimensional example is presented. The method is also applied
to empirical weather data (on a spherical surface) and brain
data (in a polar 2D domain).

\section{Methods}

\subsection{Spatio-Temporal Resonator Model}
A model for a general oscillatory phenomenon is constructed as a
superposition of several resonators (separate resonators and
their harmonics) with known angular velocities $\omega_j$
(\ie~frequencies), but unknown phases and amplitudes. These are
modeled as spatially independent realizations of stochastic
processes. The sum $\sum_{j=1}^N f_j(\vect{x},t)$ of the
oscillatory components $f_j(\vect{x},t)$ can be defined using
separate state space models. The spatially independent version
of such a resonating field can be presented as a partial
differential equation (see Eq.~\eqref{eq:oscillator}, or
\cite{Sarkka+Solin:2012} for details):
\begin{equation*}
  {\partial^2 f_j(\vect{x},t) \over \partial t^2}
    + \gamma_j {\partial f_j(\vect{x},t) \over \partial t}
    + \omega_j^2 f_j(\vect{x},t) = \xi_j(\vect{x},t),
\end{equation*}
where $\vect{x} \in \Omega$ (for some domain $\Omega \subseteq
\R^n$) denotes the spatial variable and $t \in \R_+$ represents
time. The perturbation term $\xi_j(\vect{x},t)$ is white noise, both spatially and
temporally . The above formulation also contains a
damping factor $\gamma_j$, which was assumed to be zero in
\cite{Sarkka+Solin:2012}.

Here, this formulation is extended to account
for spatial structure by assuming that the local derivative
depends not only on time, but also on surrounding locations
through some spatial linear operator. Including linear
operators that affect both the oscillation and damping results in:
\begin{equation} \label{eq:SPDE-op}
  {\partial^2 f_j(\vect{x},t) \over \partial t^2}
    + \op{A}_j {\partial f_j(\vect{x},t) \over \partial t}
    + \op{B}_j f_j(\vect{x},t) = \xi_j(\vect{x},t).
\end{equation}
This model contains three types of spatial dependency. The selection of
operators $\op{A}_j$ and $\op{B}_j$ allows the suitable definition of spatial
coupling through the first and second temporal derivative. Some spatial 
and temporal structure can also be assumed in the
process noise term $\xi_j(\vect{x},t)$ through a correlation
structure:
\begin{equation*}
  C_j(\vect{x}, \vect{x}')
    = \E[\xi_j(\vect{x},t)\xi_j(\vect{x}',t')]
    = C_{\xi,j}(\vect{x}, \vect{x}') \, \delta(t-t').
\end{equation*}

\subsection{Choosing Spatial Operators}
If the operators $\op{A}_j$ and $\op{B}_j$ are assumed to be translation
and time invariant, the corresponding Fourier
domain transfer functions $A_j(i\vectb{\nu}_x)$ and
$B_j(i\vectb{\nu}_x)$ can be calculated. Taking both spatial and temporal Fourier
transforms of Eq.~\eqref{eq:SPDE-op} results in:
\begin{multline*}
  (i\nu_t)^2 F_j(i\vectb{\nu}_x,i\nu_t)
    + (i\nu_t) A_j(i\vectb{\nu}_x) F_j(i\vectb{\nu}_x,i\nu_t) + \\
    B_j(i\vectb{\nu}_x) F_j(i\vectb{\nu}_x,i\nu_t)
    = \Xi_j(i\vectb{\nu}_x,i\nu_t).
\end{multline*}
Solving $F_j$ from above provides:
\begin{equation*}
  F_j(i\vectb{\nu}_x,i\nu_t) =
    {\Xi_j(i\vectb{\nu}_x,i\nu_t) \over (i\nu_t)^2
      + (i\nu_t)A_j(i\vectb{\nu}_x) + B_j(i\vectb{\nu}_x)},
\end{equation*}
which corresponds to the spectral density:
\begin{equation*}
  S_j(i\vectb{\nu}_x,i\nu_t) =
    {Q_j(\vectb{\nu}_x) \over \norm{(i\nu_t)^2
      + (i\nu_t)A_j(i\vectb{\nu}_x) + B_j(i\vectb{\nu}_x)}^2},
\end{equation*}
where $Q_j(\vectb{\nu}_x) = |\Xi_j(i\vectb{\nu}_x,i\nu_t)|^2$ is
the spectral density of $\xi_j$. If the operators
$\op{A}_j$ and $\op{B}_j$ are assumed to be formally Hermitian, the identities
$A_j(i\vectb{\nu}_x) = A_j(-i\vectb{\nu}_x)$ and
$B_j(i\vectb{\nu}_x) = B_j(-i\vectb{\nu}_x)$ hold, which
simplifies the spectral density to:
\begin{equation*}
  S_j(i\vectb{\nu}_x,i\nu_t) =
    {Q_j(\vectb{\nu}_x) \over \left[\nu_t^2 - B_j(i\vectb{\nu}_x)\right]^2 +\nu_t^2 A_j^2(i\vectb{\nu}_x)}.
\end{equation*}

The divisor derivative zeros of the system  suggest that the
system has a temporal resonance of $\nu_t^2 =
B_j(i\vectb{\nu}_x) - A_j^2(i\vectb{\nu}_x)/2$. The
temporal oscillation is included in the angular velocity of $\omega_j$ by
setting $B_j(i\vectb{\nu}_x) = A_j^2(i\vectb{\nu}_x)/2 +
\omega_j^2$. This gives the spectral density in the form:
\begin{equation*}
  S_j(\vectb{\nu}_x,\nu_t) =
    {Q_j(\vectb{\nu}_x) \over (\nu_t^2 - A_j^2(i\vectb{\nu}_x)/2 - \omega_j^2)^2 + \nu_t^2 A_j^2(i\vectb{\nu}_x)}.
\end{equation*}

According to Bochner's theorem \cite{DaPrato:1992} every
positive definite function is the Fourier transform of a Borel
measure. This requires the spectral density to be positive
everywhere in order to be a valid Fourier transform of a
covariance function. This condition is fulfilled if
$Q_j(\vectb{\nu}_x)$ is a positive function (\ie~a valid
spectral density). To ensure the causality and stability of the
system $A_j(i\vectb{\nu}_x)$ must be chosen such that it is a
positive function, which corresponds to the operator $\op{A}_j$
being positive (semi)definite. The operator
$\op{B}_j$ is also chosen to be positive, which results in the condition
$A_j^2(i\vectb{\nu}_x)/2 +\omega_j^2 \geq 0$. This holds, if
$A_j$ is real and positive. Zero values in the spectrum
correspond to infinite peaks. However, this does not seem to be a
problem, because if both operators are zero the model falls
back to being spatially independent, where the only spatial
structure comes from the process noise term $\xi(\vect{x},t)$.

To make the model actually useful, some choices must be made.
The coupling of $\op{A}_j$ and $\op{B}_j$ is
determined by the condition $\op{B}_j = \op{A}_j^2/2 + \omega_j^2$,
and the operator $\op{A}_j$ must be positive semidefinite.
Examples of such operators are the identity operator $\op{I}$
and the negative Laplacian $-\Delta = -\nabla^2$. Therefore,
the following operator structure is considered:
\begin{equation}
  \begin{split} \label{eq:A-and-B}
    \op{A}_j &= \gamma_j\op{I} - \chi_j\nabla^2 \\
    \op{B}_j &= {1 \over 2} (\gamma_j - \chi_j\nabla^2)^2 + \omega_j^2 \\
             &= {\gamma_j^2 \over 2} - \gamma_j\chi_j\nabla^2 + {\chi_j^2 \over 2} \nabla^4 + \omega_j^2,
  \end{split}
\end{equation}
where $\gamma_j, \chi_j \geq 0$ are some non-negative constants
and $\nabla^4$ is the so-called biharmonic operator. 

A covariance function for $\xi_j(\vect{x},t)$
must be chosen, which can be virtually any spatial stationary covariance
function 
\begin{equation*}
  \E[\xi_j(\vect{x},t)\xi_j(\vect{x}',t')]
    = C_{\xi,j}(\vect{x}-\vect{x}') \, \delta(t-t').
\end{equation*}
In theory, the covariance function $C_{\xi,j}$ could also be
non-stationary.

\subsection{Modeling Spatio-Temporal Data}
Combining all the components in the model provides the solution
as a superposition of all the oscillator components
$f(\vect{x},t) = \sum_{j=1}^N f_j(\vect{x},t)$. The oscillator
component $f_j(\vect{x},t)$ is defined by a stochastic partial
differential equation with the Dirichlet boundary conditions:
\begin{equation*}
  {\partial^2 f_j(\vect{x},t) \over \partial t^2}
    + \op{A}_j {\partial f_j(\vect{x},t) \over \partial t}
    + \op{B}_j f_j(\vect{x},t) = \xi_j(\vect{x},t),
\end{equation*}
for $(\vect{x},t) \in \Omega\times\R_+$, and $f_j(\vect{x},t) =
0$ for $(\vect{x},t) \in \partial\Omega\times\R_+$, for all
$j=1,2,\ldots,N$. The state of the system is defined as a
combination of the periodic oscillating fields and their first
temporal derivatives:
\begin{equation*}
  \vect{f}(\vect{x},t) =
  \begin{bmatrix} f_1(\vect{x},t) & \!\!{\partial \over \partial t} f_1(\vect{x},t) & \!\!\ldots & \!\!f_N(\vect{x},t) & \!\!{\partial \over \partial t} f_N(\vect{x},t) \end{bmatrix}\T\!.
\end{equation*}
This leads to the linear state space model, which can be
expressed in the following form:
\begin{equation}
  \begin{split} \label{eq:spatio-temporal-resonator}
  {\partial \vect{f}(\vect{x},t) \over \partial t}  &= \vectb{\op{F}} \, \vect{f}(\vect{x},t) + \vect{L} \, \vectb{\xi}(\vect{x},t) \\
  \vect{y}_k &= \vectb{\op{H}}_k \vect{f}(\vect{x},t_k) + \vect{r}_k,
  \end{split}
\end{equation}
where $\vectb{\op{F}}$ is a block-diagonal matrix such that each
block $j$ consists of a $2 \times 2$ matrix of linear operators,
and $\vect{L}$ is a block-diagonal matrix consisting of $2
\times 1$ blocks:
\begin{equation*}
  \vectb{\op F}_j = \begin{bmatrix} 0 & \op{I} \\ -\op{B}_j & -\op{A}_j \end{bmatrix}
     \quad \text{and} \quad
     \vect{L}_j = \begin{bmatrix} 0 \\ 1 \end{bmatrix}.
\end{equation*}
In step $k$, the observed values are $\vect{y}_k \in \R^{d_k}$. 
The measurement model is constructed by defining a functional
$\vectb{\op{H}}_k$ through which the model is observed at
discrete time steps $t_k$ at known locations
$\vect{x}^\text{obs}_i \in \Omega, i=1,2,\ldots,d_k$, that is 
$\vect{f} \mapsto \vect{f}(\vect{x}^\text{obs},t_k)$. The
measurement noise term $\vect{r}_k \sim \N(\vect{0},\vect{R}_k)$
in Eq.~\eqref{eq:spatio-temporal-resonator} is a Gaussian random
variable of dimension $d_k$. For notational convenience, the possibility 
of $\vectb{\op{F}}$ depending on time has been omitted.
However, it is included in one of the demonstrations,
where the oscillation frequencies in $\omega_j(t)$ change over
time.

\subsection{Infinite-Dimensional Kalman Filtering}
The Kalman filter (see, \eg, \cite{Sarkka:2013}) is an 
algorithm for solving the state estimation problem, which refers to 
the inverse problem of estimating the state trajectory of the stochastic process 
$\vect{f}(\vect{x},t)$ based on the noisy observations $\vect{y}_1, 
\vect{y}_2, \ldots, \vect{y}_k$. The Kalman
filter solution is the statistically optimal solution in a Bayesian 
sense given the model of the system.

Eq.~\eqref{eq:spatio-temporal-resonator} is the
infinite-dimensional counterpart of a continuous-time state
space model, where the linear matrix evolution equation has been
replaced by a linear differential operator equation
(\cf~\cite{Sarkka:AISTATS:2012}). The first equation (the
dynamic model) in \eqref{eq:spatio-temporal-resonator} is an
infinite-dimensional linear stochastic differential equation
\cite{DaPrato:1992}. Here, $\vectb{\op{F}}$ is a differential
operator, and the equation is an \emph{evolution type}
stochastic partial differential equation (SPDE)
\cite{DaPrato:1992,Chow:2007}.

Treating the temporal variable separately in the evolution type
SPDE enables the use of infinite-dimensional optimal estimation
methods. However, these methods are meant for discrete time
estimation, and therefore the evolution
equation needs to be discretized with respect to time. First, the evolution
operator is formed:
\begin{equation*}
  \vectb{\op{U}}(\Delta t) = \exp\left(\Delta t \,\vectb{\op{F}} \right),
\end{equation*}
where $\exp(\cdot)$ is the operator exponential function. A
solution to the stochastic equation can now be given as
(see \cite{DaPrato:1992,Sarkka:AISTATS:2012} for details):
\begin{multline} \label{eq:time-discretization}
  \vect{f}(\vect{x},t_{k+1}) = \vectb{\op{U}}(t_{k+1}-t_k) \, \vect{f}(\vect{x},t_k) \\
    + \int_{t_k}^{t_{k+1}} \vectb{\op{U}}(t_{k+1}-\tau) \, \vect{L} \, \vectb{\xi}(\vect{x},\tau) \dd \tau,
\end{multline}
where $t_{k+1}$ and $t_k < t_{k+1}$ are arbitrary. The second
term is a Gaussian process with covariance function
$\vect{Q}(\vect{x},\vect{x}';t_k,t_{k+1}) = \int_{t_k}^{t_{k+1}}
\vectb{\op{U}}(t_{k+1}-\tau) \, \vect{L} \,
\vect{C}_{\xi}(\vect{x},\vect{x}') \, \vect{L}\T \,
\vectb{\op{U}}^*(t_{k+1}-\tau) \dd \tau$.
This leads to the following discrete-time model:
\begin{equation} \label{eq:inf-dim-state-space-discrete}
\begin{split}
  \vect{f}(\vect{x},t_k) &= \vectb{\op{U}}(\Delta t_k) \, \vect{f}(\vect{x},t_{k-1})
    + \vect{q}_k(\vect{x}) \\
  \vect{y}_k &= \vectb{\op{H}}_k \, \vect{f}(\vect{x},t) + \vect{r}_k,
\end{split}
\end{equation}
where $\Delta t_k = t_k - t_{k-1}$ and the process noise 
$\vect{q}_k(\vect{x}) \sim \mathcal{GP}(\vect{0},
\vect{Q}(\vect{x},\vect{x}'; \Delta t_k))$. This discretization
is not an approximation, but is  the so-called \emph{mild solution}
to the infinite-dimensional differential equation
\cite{DaPrato:1992}.

\subsubsection{Infinite-Dimensional Kalman Filter and Smoother}
\label{sec:infinite-dimensional-Kalman-filter}
The \emph{infinite-dimensional Kalman filter}
\cite{Tzafestas:1978,Omatu+Seinfeld:1989,Cressie+Wikle:2002} is
a closed-form solution to the infinite-dimensional linear
filtering problem \eqref{eq:inf-dim-state-space-discrete}. 
Here, a two-step scheme is presented, which first calculates the
marginal distribution of the next step using the known system
dynamics. The following formulation uses a notation similar to
\cite{Sarkka:AISTATS:2012} and can be compared to the
finite-dimensional Kalman filter \cite{Grewal+Andrews:2001}.

The infinite-dimensional \textbf{prediction step} can be expressed
as follows:
\begin{equation}
\begin{split} \label{eq:infKalman-filter-prediction}
    \vect{m}_{k \mid k-1}(\vect{x}) &= \vectb{\op U}(\Delta t_k)\,\vect{m}_{k-1 \mid k-1}(\vect{x}) \\
    \vect{C}_{k \mid k-1}(\vect{x},\vect{x}') &= \vectb{\op U}(\Delta t_k)\,\vect{C}_{k-1 \mid k-1}(\vect{x},\vect{x}')\,\vectb{\op U}^*(\Delta t_k) \\ &\quad + \vect{Q}(\vect{x},\vect{x}';\Delta t_k),
\end{split}
\end{equation}
where $(\cdot)^*$ denotes an adjoint, which in practice swaps the
roles of inputs $\vect{x}$ and $\vect{x}'$, and operates from the
right. The operator adjoint can be seen as an operator version
of a matrix transpose.
The recursive iteration is initialized by presenting the prior information 
in the form $\vect{f}(\vect{x},t_0) \sim \mathcal{GP}\left(\vect{m}_0(\vect{x}),
\vect{C}_0(\vect{x},\vect{x}')\right)$.

The algorithm then uses each observation to update the
distribution to match the new information obtained by the
measurement in step $k$. This is the infinite-dimensional
\textbf{update step}:
\begin{equation}
\begin{split} \label{eq:infKalman-filter-update}
    \vect{S}_k &= \vectb{\op H}_k \, \vect{C}_{k \mid k-1}(\vect{x},\vect{x}') \, \vectb{\op H}_k^* + \vect{R}_k \\
    \vect{K}_k(\vect{x}) &= \vect{C}_{k \mid k-1}(\vect{x},\vect{x}') \, \vectb{\op H}_k^* \, \vect{S}_k^{-1} \\
    \vect{m}_{k \mid k}(\vect{x}) &= \vect{m}_{k \mid k-1}(\vect{x}) + \vect{K}_k(\vect{x}) \, \left(\vect{y}_k - \vectb{\op H}_k\vect{m}_{k \mid k-1}(\vect{x})\right) \\
    \vect{C}_{k \mid k}(\vect{x},\vect{x}') &= \vect{C}_{k \mid k-1}(\vect{x},\vect{x}') - \vect{K}_k(\vect{x}) \, \vect{S}_k \, \vect{K}_k^*(\vect{x}),
\end{split}
\end{equation}
where $(\cdot)^{-1}$ denotes the matrix inverse. As a result, the
filtered forward-time posterior process in step $k$ (time~$t_k$)
is given by $\vect{f}_{k \mid k}(\vect{x}) \sim
\mathcal{GP}\left(\vect{m}_{k\mid k}(\vect{x}),\,
\vect{C}_{k\mid k}(\vect{x},\vect{x}')\right)$.

The purpose of optimal (fixed-interval) smoothing is to obtain
in closed-form the marginal posterior distribution of the state
$\vect{f}_k$ in time step $t_k$, which is conditional on all the
measurements $\vect{y}_{1:T}$, where $k \in [1,\ldots,T]$ is a
fixed interval.

The infinite-dimensional Rauch--Tung--Striebel (RTS) smoother
equations are written so that they utilize the Kalman filtering
results $\vect{m}_{k \mid k}(\vect{x})$ and $\vect{C}_{k \mid
k}(\vect{x},\vect{x}')$ as a forward sweep, and then perform a
backward sweep to update the estimates to match the forthcoming
observations. The smoother's backward sweep may be expressed with
the following infinite-dimensional \textbf{RTS smoothing
equations} \cite{Sarkka:AISTATS:2012}:
\begin{align} 
    \vect{m}_{k+1 \mid k}(\vect{x}) &= \vectb{\op U}(\Delta t_k) \, \vect{m}_{k \mid k}(\vect{x}) \nonumber \\
    \vect{C}_{k+1 \mid k}(\vect{x},\vect{x}') &= \vectb{\op U}(\Delta t_k) \, \vect{C}_{k \mid k}(\vect{x},\vect{x}') \, \vectb{\op U}^*(\Delta t_k) \nonumber\\
    &\quad + \vect{Q}_{k}(\vect{x},\vect{x}'; \Delta t_k) \nonumber\\
    \vectb{\op G}_k &= \vect{C}_{k \mid k}(\vect{x},\vect{x}') \, \vectb{\op U}^*(\Delta t_k) \, \left[\vect{C}_{k+1 \mid k}(\vect{x},\vect{x}')\right]^{-1} \nonumber\\
    \vect{m}_{k \mid T}(\vect{x}) &= \vect{m}_{k \mid k}(\vect{x}) \nonumber\\
    &\quad + \vectb{\op G}_k \, \left[\vect{m}_{k+1 \mid T}(\vect{x}) - \vect{m}_{k+1 \mid k}(\vect{x})\right] \nonumber\\
    \vect{C}_{k \mid T}(\vect{x},\vect{x}') &= \vect{C}_{k \mid k}(\vect{x},\vect{x}') + \vectb{\op G}_k \, \big(\vect{C}_{k+1 \mid T}(\vect{x},\vect{x}') \nonumber\\
    & \quad - \vect{C}_{k+1 \mid k}(\vect{x},\vect{x}')\big)\vectb{\op G}_k^*. \nonumber\\
\label{eq:infRTS}
\end{align} 

In the above equations, $(\cdot)^{-1}$ denotes the operator inverse. In addition, note
that $\vectb{\op G}_k$ is a linear operator whose kernel is
defined via the covariance kernels of the filtering results. This
makes the notation slightly more challenging.

Once both the Kalman filtering and
Rauch--Tung--Striebel sweeps are performed on the model given the observed
data, the marginal posterior is obtained, which can be represented as the
Gaussian process:
\begin{equation*}
  \vect{f}(\vect{x},t_k \mid \vect{y}_{1:T}) \sim
       \mathcal{GP}\left( \vect{m}_{k \mid T}(\vect{x}),
          \vect{C}_{k \mid T}(\vect{x},\vect{x}') \right),
\end{equation*}
where the observed values $\vect{y}_k \in \R^{d_k}$ are given at
discrete time points $t_k, k=1,2,\ldots,T$, and measured at
known locations $\vect{x}^\text{obs}_{i,k} \in \Omega,
i=1,\ldots,d_k$. The resulting process functions can be
evaluated at any test point $\vect{x}_* \in \Omega$ by simply
considering an appropriate measurement functional $\mathcal{H}$.
Thus, the marginal posterior of the value of
$\vect{f}(\vect{x}_*,t_k)$ in $\vect{x}_*$ at time instant $t_k$
is:
\begin{multline*}
  p\left(\vect{f}(\vect{x}_*,t_k) \mid \vect{y}_{1:T}\right) = \\
   \N\left(\vect{f}(\vect{x}_*,t_k) \mid
     \vect{m}_{k \mid T}(\vect{x}_*), \vect{C}_{k \mid T}(\vect{x}_*,\vect{x}_*)\right).
\end{multline*}

Values could also be predicted at more time steps. A
test time point $t_*$ should be taken into account  when performing
the time discretization. The state of the system
$\vect{f}(\vect{x},t_*)$ would be predicted in this step, but because
there are no data, no updating step is needed.

One detail worthy of note is the connection between the
standard (in this case) spatial GP model and the evolution type
state space SPDE. If the temporal evolution model is left out,
that is $\vectb{\op F} = 0$ and $\vect{Q}_c(\vect{x},
\vect{x}')=0$ are chosen, the estimation task for a spatial GP model could
be solved by considering only one measurement step and using the
same equations.

\subsubsection{Hilbert Space Methods}
The infinite-dimensional Kalman filtering and smoothing
equations can be converted to a tractable form by either
finite difference approximations or introducing
Hilbert space methods (basis function approximations). 
The eigenfunction expansion of the linear operator is considered and
combined with the infinite-dimensional framework. By
truncating the expansion, a finite-dimensional
approximate solution is obtained, which can be evaluated.

In the spatio-temporal resonator model, the
negative Laplace operator can be considered in some spatial domain $\Omega$
subject to Dirichlet boundary conditions. This results in an
eigenfunction equation in the form $-\nabla^2 \psi_n(\vect{x}) =
\lambda_n \psi_n(\vect{x})$, where $\psi_n(\vect{x})$ is an
eigenfunction and $\lambda_n$ is the corresponding eigenvalue for
each $n = 1,2,\ldots,N$ and spatial coordinate $\vect{x} \in
\Omega$. The solution presented here $\vect{f}(\vect{x},t)$ will be transformed
to a new basis, which is given by the eigendecomposition of the
linear operator in $\Omega$. This new basis decodes the spatial
structure so that only $\vect{\tilde f}(t)$ remains, being a
finite-dimensional approximation of $\vect{f}(\vect{x},t)$.

After the time-discretization step in
\eqref{eq:time-discretization}, the finite-dimensional
approximation of the $s$ component model leads to a
$sN$-dimensional state space model:
\begin{equation*}
  \begin{split}
    \vect{\tilde f}_{k+1} &= \vect{A}_k \, \vect{\tilde f}_k + \vect{q}_k \\
    \vect{y}_k &= \vect{H}_k \, \vect{\tilde f}_k + \vect{r}_k,
  \end{split}
\end{equation*}
where the sparse dynamic model $\vect{A}_k$ and the noise term
$\vect{q}_k \sim \N(\vect{0},\vect{Q}_k)$ are given in the basis
defined by the eigenfunction expansion of $\vectb{\op F}$ in
$\Omega$ (see the supplementary material in
\cite{Sarkka:AISTATS:2012} for detailed equations).
The explicit form of $\vect{H}_k$ is determined by
the basis functions evaluated at the observation locations 
$\vect{x}^\text{obs}$ at time $t_k$.

\subsection{Parameter Estimation}
All the parameters needed in the model (typically
the unknowns in the covariance function, the damping constants,
and the measurement noise variance) are summarized as a vector quantity
$\vectb{\theta}$. For notational convenience, the parameters have
not been explicitly written out in
\eqref{eq:infKalman-filter-prediction} and
\eqref{eq:infKalman-filter-update}, but could be included as
$\vectb{\op U}(\Delta t, \vectb{\theta})$,
$\vect{Q}(\vect{x},\vect{x}';t,t'; \vectb{\theta})$ and
$\vect{R}_k(\vectb{\theta})$. Based on these results, 
the marginal likelihood of the measurements can be computed, given
$\vectb{\theta}$ (see, \eg, \cite{Singer:2011}):
\begin{equation*}
  p(\vect{y}_1, \ldots, \vect{y}_T \mid \vectb{\theta}) =
    \prod_{k=1}^T \N\left(\vect{y}_k \mid
    \vectb{\op H}_k \, \vect{m}_{k \mid k-1}(\vect{x},\vectb{\theta}),
    \vect{S}_k(\vectb{\theta})\right).
\end{equation*}
Hence, the marginal log-likelihood function for maximum
likelihood (ML) estimation can be given with the help of
the predicted mean $\vect{m}_{k \mid k-1}(\vect{x},\vectb{\theta})$ and the
innovation covariance $\vect{S}_k(\vectb{\theta})$ in \eqref{eq:infKalman-filter-prediction} and \eqref{eq:infKalman-filter-update}:
\begin{multline} \label{eq:log-likelihood}
  \hspace{-8pt}%
  \ell (\vectb{\theta}) =
    -\frac{1}{2} \sum_{k=1}^T\! \log |2\pi \vect{S}_k(\vectb{\theta})|
    -\frac{1}{2} \sum_{k=1}^T\!
      \left( \vect{y}_k\! -\! \vect{\op H}_k\,\vect{m}_{k \mid k-1}(\vect{x},\vectb{\theta}) \right)\T \\ \times
      \vect{S}_k^{-1}(\vectb{\theta})
      \left( \vect{y}_k\! -\! \vectb{\op H}_k\,\vect{m}_{k \mid k-1}(\vect{x},\vectb{\theta}) \right).
\end{multline}
The parameter estimation problem now reverts to the maximization
of the log-likelihood function: $\hat{\vectb{\theta}} =
\arg\,\max_{\vectb{\theta}} \, \ell(\vectb{\theta})$.

Using the log-likelihood function, the un-normalized posterior distribution
 could also be formed easily, which would allow the
computation of maximum a posteriori (MAP) estimates or the use of a
Metropolis--Hastings type of Markov chain Monte Carlo (MCMC) for
integration over the parameters.

\begin{figure}[t]
  \centering
  \includegraphics[width=0.5\textwidth]{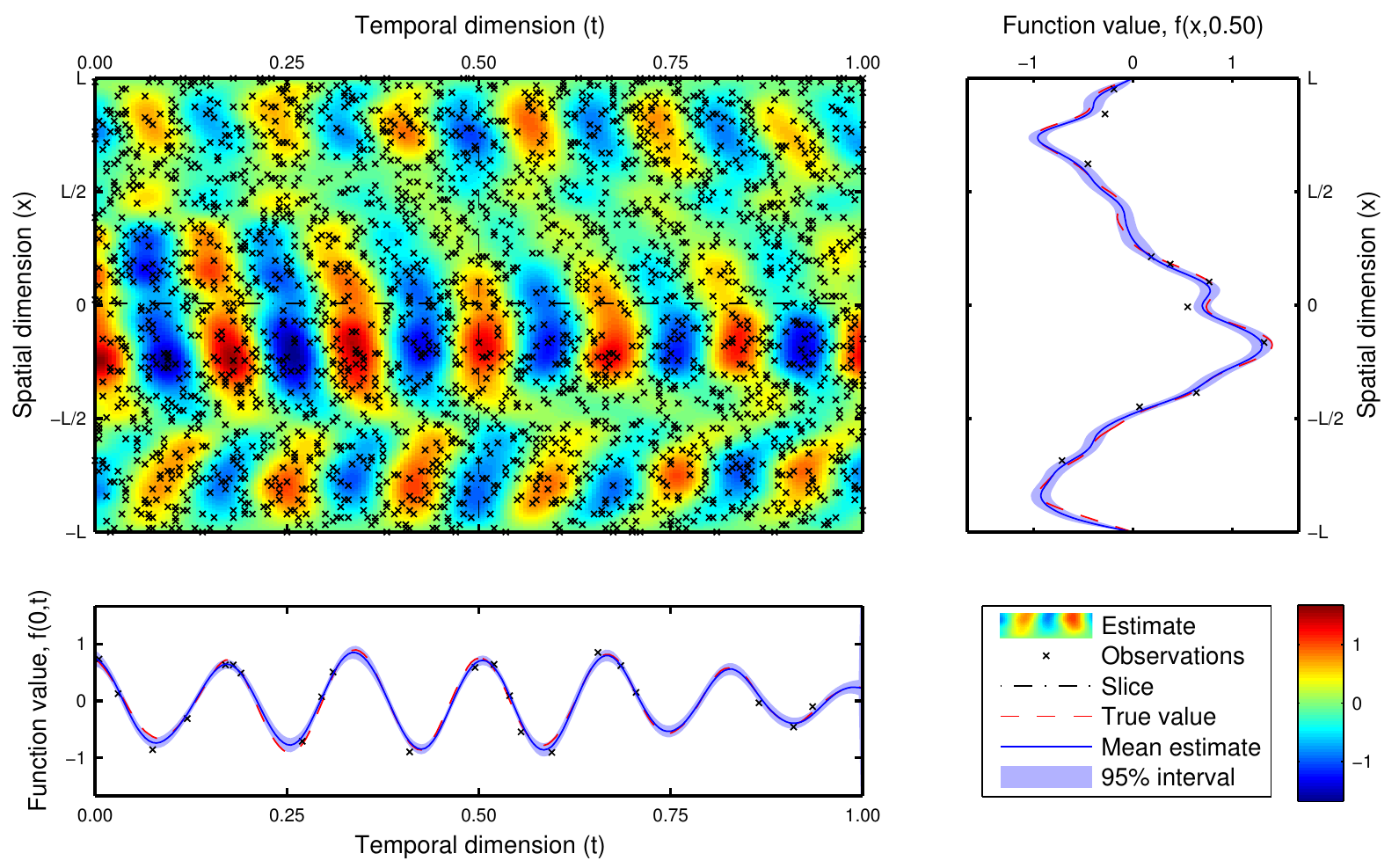}
  \caption{This figure illustrates an example of a stochastic
    resonator realization in one spatial dimension. The left-hand figure
    shows the observation locations and the corresponding estimate of
    the oscillating field. The state at $t=0.50\,$s is shown in the
    right-hand figure, where the estimate is presented together with the dashed
    true values of the process.}
  \label{fig:simulation-in-1D}
\end{figure}

\begin{figure*}[!t]
  \centering
  \subfloat[Slow temperature bias]{%
    \includegraphics[width=0.40\textwidth]{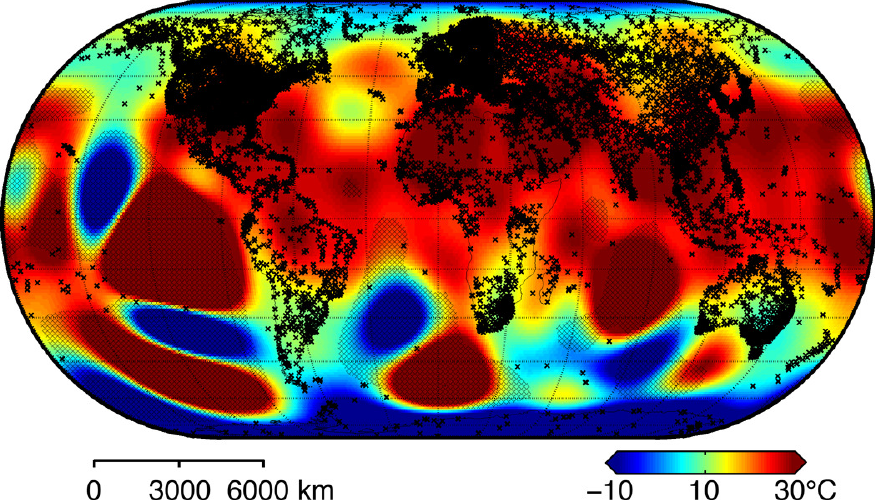}%
    \label{fig:temperature-results-a}
  }%
  \hspace{0.1\textwidth}%
  \subfloat[Oscillatory structure with two harmonics]{%
    \includegraphics[width=0.40\textwidth]{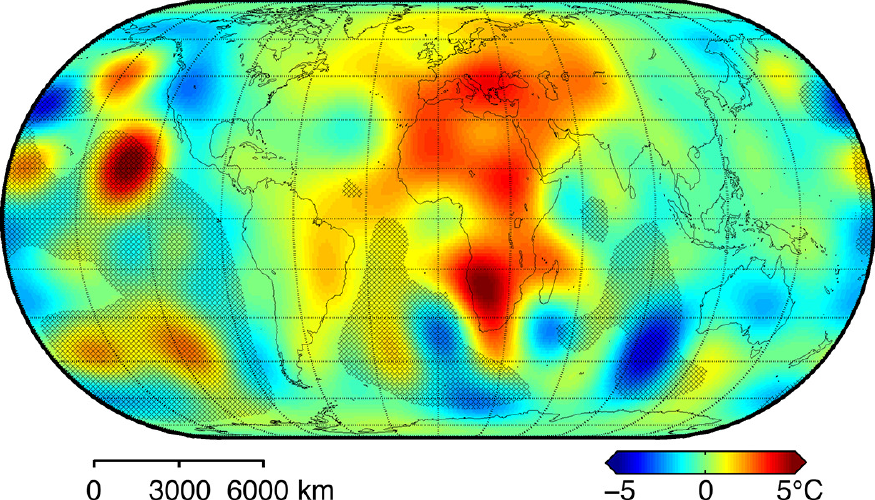}%
    \label{fig:temperature-results-b}
  }%
  \caption{Slow bias and resonator maps for temperatures as a
    snapshot on July~8, 2011 at 2~PM (GMT). The weather stations
    are marked with crosses and areas of uncertainty are hatched.}
  \label{fig:temperature-results}
\end{figure*}

\section{Results}

\subsection{Illustrative One-Dimensional Example}
\label{sec:demo-simple}
Visualizations for simulated data in one spatial dimension are
shown in Fig.~\ref{fig:simulation-in-1D} as an example of the spatio-temporal resonator model. 
The model
contains one resonator $f(x,t)$ oscillating at a frequency of
$6\,$Hz, where $x \in [-L,L]$ over a time-span $t \in [0,1]$
(in seconds).

The Mat{\'e}rn covariance function is considered (see, \eg,
\cite{Rasmussen:2006}) for the perturbing dynamic noise. This
class of stationary isotropic covariance functions is widely
used in many applications and their parameters have understandable
interpretations. A Mat{\'e}rn covariance function can be expressed
as:
\begin{equation} \label{eq:Matern-covariance}
  C(r) = s^2\,\frac{2^{1-\nu}}{\Gamma(\nu)}\left(\sqrt{2\nu}\,\frac{r}{l}\right)^\nu K_\nu\!\left(\sqrt{2\nu}\,\frac{r}{l}\right),
\end{equation}
where $r = \norm{\vect{x}-\vect{x}'}$, $\Gamma(\cdot)$ is the
Gamma function and $K_\nu(\cdot)$ is the modified Bessel
function. The covariance function is characterized by three
parameters: a smoothness parameter $\nu$, a distance scale
parameter $l$, and a strength (magnitude) parameter $\sigma$, all
of which are positive.

For simulating data, the model parameters were chosen so that
$\gamma = 1$ and $\chi = 0.01$. The perturbing dynamic noise
covariance function parameters were: $\nu=3/2$,
$l=0.1\,L$, and $s = 25$. The Gaussian measurement noise
variance was $\sigma^2 = 0.1^2$. A truncated
eigenfunction decomposition with 32 eigenfunctions was used. Altogether,
2500 noisy observations were considered.

The model parameters are fitted by optimizing the marginal
log-likelihood \eqref{eq:log-likelihood} using a conjugate
gradient optimizer and square root versions of the filtering
equations
\eqref{eq:infKalman-filter-prediction}--\eqref{eq:infRTS} for
numerical stability. To avoid bad local minima, ten random
restarts were attempted, and the run with the best marginal
likelihood was selected. The optimized parameters were $\hat{\sigma}^2
\approx 0.098^2$, $\hat{l} \approx 0.106\,L$, $\hat{s} \approx
30.8$, $\hat{\gamma} \approx 0.690$, and $\hat{\chi} \approx
0.015$.

%

Fig.~\ref{fig:simulation-in-1D} shows the estimation mean
$m(x,t)$ of the true space--time oscillating field $f(x,t)$ as a
color surface plot. The measurement locations in space--time are
shown as crosses. The oscillatory behavior is clear, and
even clearer along the slice $x=0$, which is shown in the lower
figure. A slice $f(x,0.5)$ along the spatial dimension is shown
on the right. Both slices contain the true values for $f$
(dashed), the estimate mean (solid), and a shaded 95\%
uncertainty interval.

\begin{figure*}
  \centering
  \subfloat[Cardiac noise amplitude]{%
     \includegraphics[width=0.4\textwidth]{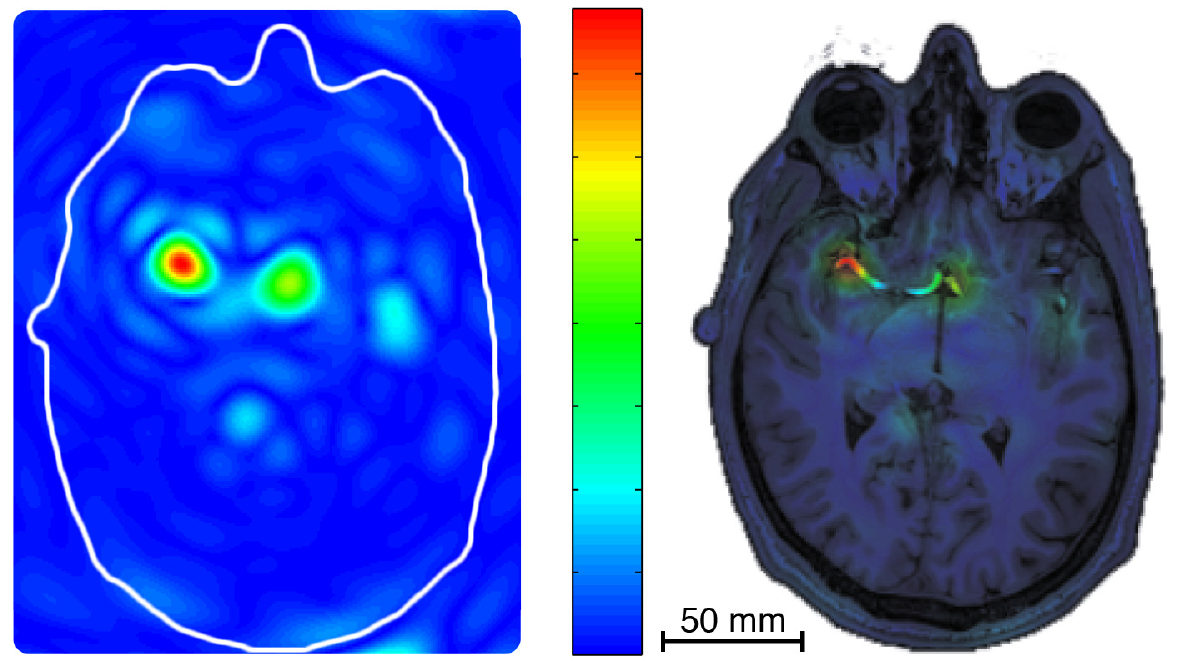}%
     \label{fig:card-amp}}%
  \hspace{0.1\textwidth}%
  \subfloat[Respiratory noise amplitude]{%
     \includegraphics[width=0.4\textwidth]{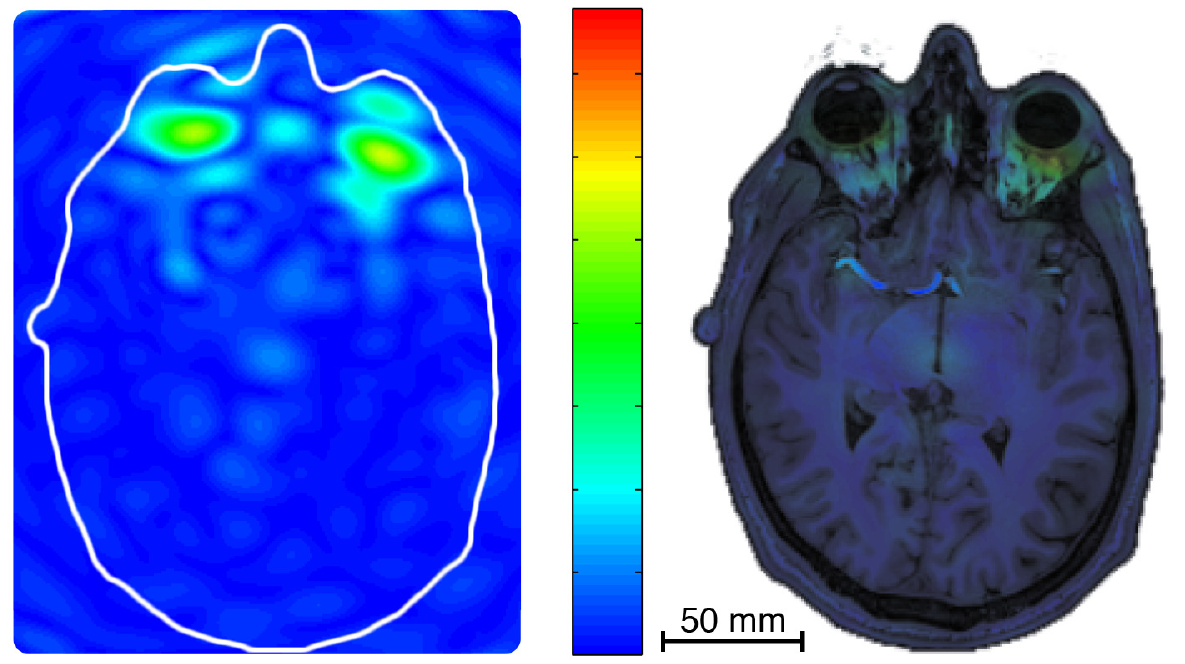}%
     \label{fig:resp-amp}}%
  \caption{Mean amplitude maps for the physiological noise components.
    The results are shown both in isolation and overlaid on top of the
    corresponding anatomical image.}
  \label{fig:amplitude-map-2d}
\end{figure*}

\subsection{Weather Data on the Surface of the Globe}
\label{sec:demo-weather}
In this section, the proposed methods are applied to hourly observations of
temperature readings in centigrade, which were collected
worldwide by \emph{National Environmental Satellite, Data, and
Information Services} (NESDIS)\footnote{The dataset is available
for download from U.S. National Climatic Data Center:
\url{http://www7.ncdc.noaa.gov/CDO/cdoselect.cmd} (accessed
December 12, 2011).}. A subset of the data was considered 
consisting of hourly temperature measurements for one month (July,
2011), resulting in a time series of $745$ temporal  points. The
temperatures were recorded at $11\,344$ different spatial 
locations, for which the longitudinal and latitudinal coordinates
are known and assumed exact. The locations of the stations
are marked as crosses in
Fig.~\ref{fig:temperature-results-a}. However, not all
stations provide hourly data; altogether there are $5\,637\,501$ measurements.

A model with three latent components $f_j(\vect{x},t)$ was used.
The first latent component, a bias term with exponentially decaying
memory, accounts for the slow drifting of the mean temperature, and 
the other two are oscillatory components for the daily variation of 
temperature. The first of these two oscillates at the constant base 
frequency of $1/$day, and the second is the first harmonic (frequency 
$2$/day). Similarly as in \cite{Sarkka+Solin:2012}, the bias term
was constructed as an oscillator with zero frequency. This corresponds
to a spatio-temporal Wiener velocity model.
The spatial covariance function of the process noise was fixed to the
squared-exponential covariance function.

The model parameters (covariance function magnitudes, length
scales, and the Gaussian measurement noise variance) were all
optimized with respect to marginal likelihood using a few random
restarts to avoid bad local minima. The damping parameters were
virtually zero in all runs, therefore they were fixed to zero in
the final model. This means that the spatial dependencies stem
from the perturbation structure alone.

The estimation results for the temperature oscillation model are
presented here as a snapshot of the temperature map over the
globe on July~8, 2011 at 2~PM (GMT). The results in
Fig.~\ref{fig:temperature-results} are split into two in
order to show the influence of the slow-moving bias and the
resonating part. Fig.~\ref{fig:temperature-results-a}
also shows the spatial locations of the weather stations. The
hatched regions indicate uncertainty (standard deviation $>
2\,^\circ$C), which corresponds to the regions with very few or
no observations.

This test setup is subject to many simplifications and
assumptions, which affect the results; the surface of the Earth is
actually not a symmetrical sphere and the evident
fact that the fluctuation covariance structure is not
stationary is disregarded. However, the setup clearly 
captures two effects: the summer
on the northern hemisphere and the day--night variation
(afternoon in Europe and Africa in the figure).

\subsection{Modeling Pulsations in the Brain}
\label{sec:demo-brain}
Recent advances in \emph{functional magnetic resonance imaging}
(fMRI) techniques have demonstrated the importance of computational methods
in modeling brain data. In \cite{Sarkka+Solin:2012}, it
was shown that eliminating oscillating physiological noise
components in fMRI can be achieved by using a spatially independent
resonator model combined with Kalman filtering. This method was
named DRIFTER. This approach is now extended by showing how
to account also for spatial dependencies.

One $\sim$30$\,$s run of empirical fMRI data was considered. 
The set of fast-sampled data of one slice is used here to demonstrate the
spatio-temporal resonator model in two-dimensional polar
coordinates. This fMRI data, together with anatomical images
for one volunteer, were obtained using a $3\,$T scanner (Siemens
Skyra) located at the \emph{Advanced Magnetic Imaging Centre}
(AMI) of Aalto University School of Science using a 32-channel
receive-only head coil. For the functional imaging, the major
parameters were repetition time (TR) $77\,$ms, echo time (TE)
$20\,$ms, flip angle (FA) 60$^{\circ}$, field-of-view (FOV)
$224\,$mm, and matrix size 64$\times$64. The measurements were
performed as part of AMI Centre's local technical methods development
research and conformed to the guidelines of the Declaration of
Helsinki. The research was approved by the ethical committee in
the Hospital District of Helsinki and Uusimaa.

In order to simulate resting state conditions, the stimulus
was a fixed dot in the center of the visual field of the
volunteer. The heart and respiratory signals were recorded,
and time-locked to the fMRI data during the run. The oscillation
frequency of the physiological noise components was quasi-periodic (rather than exactly periodic), which implies that the frequencies
change over time. External reference signals with the
\emph{interacting multiple model} (IMM) approach, presented in
\cite{Sarkka+Solin:2012}, were used to estimate the frequency time series
of heart beats and respiration cycles. The cardiac frequency
alternated between $54$--$64$ beats per minute and the respiratory frequency
between $6$--$15$ cycles per minute.

One slice of fast-sampled fMRI data was used. The sampling
interval was $0.077\,$s and the whole 64$\times$64
matrix was observed during each of 350 time steps. The spatial domain $\Omega$
was chosen to be a circular disk with a radius of
$\approx155\,$mm. The spatio-temporal resonator model has three
components: a slowly moving brain blood-oxygen-level-dependent
(BOLD) signal---which also includes scanner drift and other slow
phenomena---modeled with a spatio-temporal Wiener velocity model
(see Sec.~\ref{sec:demo-weather}), and two space--time
resonators oscillating at the time-dependent cardiac and
respiratory frequencies, respectively. Here, only 
the resonators for the base frequencies were included; more complex signals
could be accounted for by including harmonics. 
An eigenfunction decomposition of the linear
operators in $\Omega$ with 300 eigenfunctions was used for each of the three components. 
Model parameters
were chosen by studying the spatially independent model first.

Figure~\ref{fig:amplitude-map-2d} shows the spatial
amplitude of physiological noise contribution averaged over
time. The results resemble the maps in
\cite{Sarkka+Solin:2012b}, where the oscillators were assumed to be spatially independent and only the final results were
spatially smoothed. This suggests that the method is able to
capture the space--time structure of the oscillations. The
cardiac influence is strong near the large cerebral arteries (see
Fig.~\ref{fig:card-amp}), and the respiration causes
artifacts near the eyes that are partly induced by movement.

\section{Conclusion and Discussion}
This paper proposed a computationally effective
stochastic partial differential equation model and an inference
method for detecting and modeling latent oscillatory structures
in spatio-temporal data. It showed how a physical
first-principles SPDE model for spatio-temporal oscillations can
be constructed, and how the Bayesian inference can be effectively
applied using infinite-dimensional Kalman filtering and Hilbert
space methods. This filtering is related to Gaussian process
regression and Hilbert space valued stochastic processes. The
proposed method allows a reduction of the complexity of a
direct GP solution from cubic to linear with respect to measurements in the
temporal dimension.

A truncated eigenfunction expansion of the Laplace
operator was used to form a finite-dimensional basis over the spatial
domain, which made it possible to revert to the traditional Kalman
filtering scheme. The eigenfunction expansions of the Laplace
operator in both spherical and Cartesian coordinates were used
in the numerical computations.

The numerical results show that the truncated expansion puts
some restrictions on the spatial short-scale variability. The
basis function approach tends to make the model spatially
smooth, a problem that has been dealt with before in many ways under the
GP regression scheme (see, \eg, \cite{Rasmussen:2006}).
However, in many applications such as in functional brain data
analysis, this is not a problem since a few hundred basis
functions are sufficient to match the required spatial resolution.

Several methodological extensions could be considered: 
In temporal sense, a broad family of quasi-periodic
oscillations can be modeled by including a sufficient number of
harmonics. Non-linear oscillator models can also be approximated,
if the realizations are periodic. This applies, for example, to
relaxation oscillators such as systems generated by the Van der
Pol oscillator. Other properties of the stochastic
oscillator model \cite{Sarkka+Solin:2012} also apply to this 
spatial extension of it. In Sec.~\ref{sec:demo-brain} the oscillation
frequency was time-dependent, and it would be possible to extend
the model to account for different regions of the spatially
extended system oscillating at different frequencies. Including
non-stationary covariance functions in the process noise term
would provide many extensions to the perturbation model
structure. Spatially, relaxing the coupling between operators
$\op{A}$ and $\op{B}$ allows various spatio-temporal models to
be accounted for. For example, if we consider in
Eq.~\eqref{eq:A-and-B} $\gamma_j = \chi_j = 0$, but
$\gamma_j\chi_j > 0$, the model corresponds to an undamped
oscillating field, where the perturbations follow wave equation
dynamics. The inference scheme is also compatible with this type
of models, which can be useful, for example, in modeling
of epidemic spread.

In general, the spatio-temporal model can mitigate the problems
related to slow sampling rates, because the spatial information can
compensate for missing temporal data. This turns
such models into powerful tools for signal reconstruction and noise
elimination, for example in fMRI studies, especially as the
computational complexity grows only linearly with the length of
the measurement session.

\section*{Acknowledgments}
The authors would like to thank Simo Vanni, Ari Laiho, and Toni
Auranen for their help with data acquisition, and Aki Vehtari and
Fa-Hsuan Lin for some helpful discussions of the topic. We also
acknowledge the computational resources provided by the Aalto
Science-IT project.

\bibliography{references}

\begin{thebibliography}{24}%
\makeatletter
\providecommand \@ifxundefined [1]{%
 \@ifx{#1\undefined}
}%
\providecommand \@ifnum [1]{%
 \ifnum #1\expandafter \@firstoftwo
 \else \expandafter \@secondoftwo
 \fi
}%
\providecommand \@ifx [1]{%
 \ifx #1\expandafter \@firstoftwo
 \else \expandafter \@secondoftwo
 \fi
}%
\providecommand \natexlab [1]{#1}%
\providecommand \enquote  [1]{``#1''}%
\providecommand \bibnamefont  [1]{#1}%
\providecommand \bibfnamefont [1]{#1}%
\providecommand \citenamefont [1]{#1}%
\providecommand \href@noop [0]{\@secondoftwo}%
\providecommand \href [0]{\begingroup \@sanitize@url \@href}%
\providecommand \@href[1]{\@@startlink{#1}\@@href}%
\providecommand \@@href[1]{\endgroup#1\@@endlink}%
\providecommand \@sanitize@url [0]{\catcode `\\12\catcode `\$12\catcode
  `\&12\catcode `\#12\catcode `\^12\catcode `\_12\catcode `\%12\relax}%
\providecommand \@@startlink[1]{}%
\providecommand \@@endlink[0]{}%
\providecommand \url  [0]{\begingroup\@sanitize@url \@url }%
\providecommand \@url [1]{\endgroup\@href {#1}{\urlprefix }}%
\providecommand \urlprefix  [0]{URL }%
\providecommand \Eprint [0]{\href }%
\providecommand \doibase [0]{http://dx.doi.org/}%
\providecommand \selectlanguage [0]{\@gobble}%
\providecommand \bibinfo  [0]{\@secondoftwo}%
\providecommand \bibfield  [0]{\@secondoftwo}%
\providecommand \translation [1]{[#1]}%
\providecommand \BibitemOpen [0]{}%
\providecommand \bibitemStop [0]{}%
\providecommand \bibitemNoStop [0]{.\EOS\space}%
\providecommand \EOS [0]{\spacefactor3000\relax}%
\providecommand \BibitemShut  [1]{\csname bibitem#1\endcsname}%
\let\auto@bib@innerbib\@empty
\bibitem [{\citenamefont {Rotermund}\ \emph {et~al.}(1990)\citenamefont
  {Rotermund}, \citenamefont {Engel}, \citenamefont {Kordesch},\ and\
  \citenamefont {Ertl}}]{Rotermund:1990}%
  \BibitemOpen
  \bibfield  {author} {\bibinfo {author} {\bibfnamefont {H.}~\bibnamefont
  {Rotermund}}, \bibinfo {author} {\bibfnamefont {W.}~\bibnamefont {Engel}},
  \bibinfo {author} {\bibfnamefont {M.}~\bibnamefont {Kordesch}}, \ and\
  \bibinfo {author} {\bibfnamefont {G.}~\bibnamefont {Ertl}},\ }\href@noop {}
  {\bibfield  {journal} {\bibinfo  {journal} {Nature}\ }\textbf {\bibinfo
  {volume} {343}},\ \bibinfo {pages} {355} (\bibinfo {year}
  {1990})}\BibitemShut {NoStop}%
\bibitem [{\citenamefont {Rzhanov}\ \emph {et~al.}(1993)\citenamefont
  {Rzhanov}, \citenamefont {Richardson}, \citenamefont {Hagberg},\ and\
  \citenamefont {Moloney}}]{Rzhanov:1993}%
  \BibitemOpen
  \bibfield  {author} {\bibinfo {author} {\bibfnamefont {Y.~A.}\ \bibnamefont
  {Rzhanov}}, \bibinfo {author} {\bibfnamefont {H.}~\bibnamefont {Richardson}},
  \bibinfo {author} {\bibfnamefont {A.~A.}\ \bibnamefont {Hagberg}}, \ and\
  \bibinfo {author} {\bibfnamefont {J.~V.}\ \bibnamefont {Moloney}},\
  }\href@noop {} {\bibfield  {journal} {\bibinfo  {journal} {Physical Review
  A}\ }\textbf {\bibinfo {volume} {47}},\ \bibinfo {pages} {1480} (\bibinfo
  {year} {1993})}\BibitemShut {NoStop}%
\bibitem [{\citenamefont {Singh}\ and\ \citenamefont
  {Sinha}(2013)}]{Singh:2013}%
  \BibitemOpen
  \bibfield  {author} {\bibinfo {author} {\bibfnamefont {R.}~\bibnamefont
  {Singh}}\ and\ \bibinfo {author} {\bibfnamefont {S.}~\bibnamefont {Sinha}},\
  }\href@noop {} {\bibfield  {journal} {\bibinfo  {journal} {Physical Review
  E}\ }\textbf {\bibinfo {volume} {87}},\ \bibinfo {pages} {012907} (\bibinfo
  {year} {2013})}\BibitemShut {NoStop}%
\bibitem [{\citenamefont {S\"arkk\"a}\ \emph
  {et~al.}(2012{\natexlab{a}})\citenamefont {S\"arkk\"a}, \citenamefont
  {Solin}, \citenamefont {Nummenmaa}, \citenamefont {Vehtari}, \citenamefont
  {Auranen}, \citenamefont {Vanni},\ and\ \citenamefont
  {Lin}}]{Sarkka+Solin:2012}%
  \BibitemOpen
  \bibfield  {author} {\bibinfo {author} {\bibfnamefont {S.}~\bibnamefont
  {S\"arkk\"a}}, \bibinfo {author} {\bibfnamefont {A.}~\bibnamefont {Solin}},
  \bibinfo {author} {\bibfnamefont {A.}~\bibnamefont {Nummenmaa}}, \bibinfo
  {author} {\bibfnamefont {A.}~\bibnamefont {Vehtari}}, \bibinfo {author}
  {\bibfnamefont {T.}~\bibnamefont {Auranen}}, \bibinfo {author} {\bibfnamefont
  {S.}~\bibnamefont {Vanni}}, \ and\ \bibinfo {author} {\bibfnamefont {F.-H.}\
  \bibnamefont {Lin}},\ }\href@noop {} {\bibfield  {journal} {\bibinfo
  {journal} {NeuroImage}\ }\textbf {\bibinfo {volume} {60}},\ \bibinfo {pages}
  {1517} (\bibinfo {year} {2012}{\natexlab{a}})}\BibitemShut {NoStop}%
\bibitem [{\citenamefont {S\"arkk\"a}\ \emph
  {et~al.}(2012{\natexlab{b}})\citenamefont {S\"arkk\"a}, \citenamefont
  {Solin}, \citenamefont {Nummenmaa}, \citenamefont {Vehtari}, \citenamefont
  {Auranen}, \citenamefont {Vanni},\ and\ \citenamefont
  {Lin}}]{Sarkka+Solin:2012b}%
  \BibitemOpen
  \bibfield  {author} {\bibinfo {author} {\bibfnamefont {S.}~\bibnamefont
  {S\"arkk\"a}}, \bibinfo {author} {\bibfnamefont {A.}~\bibnamefont {Solin}},
  \bibinfo {author} {\bibfnamefont {A.}~\bibnamefont {Nummenmaa}}, \bibinfo
  {author} {\bibfnamefont {A.}~\bibnamefont {Vehtari}}, \bibinfo {author}
  {\bibfnamefont {T.}~\bibnamefont {Auranen}}, \bibinfo {author} {\bibfnamefont
  {S.}~\bibnamefont {Vanni}}, \ and\ \bibinfo {author} {\bibfnamefont {F.-H.}\
  \bibnamefont {Lin}},\ }in\ \href@noop {} {\emph {\bibinfo {booktitle}
  {Proceedings of {ISMRM} 2012}}},\ \bibinfo {series and number} {\bibinfo
  {number} {1535}}\ (\bibinfo {organization} {The International Society for
  Magnetic Resonance in Medicine},\ \bibinfo {year} {2012})\BibitemShut
  {NoStop}%
\bibitem [{\citenamefont {Burrage}\ \emph {et~al.}(2008)\citenamefont
  {Burrage}, \citenamefont {Lenane},\ and\ \citenamefont
  {Lythe}}]{Burrage+Lenane+Lythe:2008}%
  \BibitemOpen
  \bibfield  {author} {\bibinfo {author} {\bibfnamefont {K.}~\bibnamefont
  {Burrage}}, \bibinfo {author} {\bibfnamefont {I.}~\bibnamefont {Lenane}}, \
  and\ \bibinfo {author} {\bibfnamefont {G.}~\bibnamefont {Lythe}},\
  }\href@noop {} {\bibfield  {journal} {\bibinfo  {journal} {{SIAM} journal on
  scientific computing}\ }\textbf {\bibinfo {volume} {29}},\ \bibinfo {pages}
  {245} (\bibinfo {year} {2008})}\BibitemShut {NoStop}%
\bibitem [{\citenamefont {Hartikainen}\ \emph {et~al.}(2012)\citenamefont
  {Hartikainen}, \citenamefont {Sepp\"anen},\ and\ \citenamefont
  {S\"arkk\"a}}]{Hartikainen+Seppanen+Sarkka:2012}%
  \BibitemOpen
  \bibfield  {author} {\bibinfo {author} {\bibfnamefont {J.}~\bibnamefont
  {Hartikainen}}, \bibinfo {author} {\bibfnamefont {M.}~\bibnamefont
  {Sepp\"anen}}, \ and\ \bibinfo {author} {\bibfnamefont {S.}~\bibnamefont
  {S\"arkk\"a}},\ }in\ \href@noop {} {\emph {\bibinfo {booktitle} {In
  Proceedings of the 29th {International Conference on Machine Learning}
  ({ICML} 2012)}}}\ (\bibinfo {year} {2012})\BibitemShut {NoStop}%
\bibitem [{\citenamefont {Hale}(1997)}]{Hale:1997}%
  \BibitemOpen
  \bibfield  {author} {\bibinfo {author} {\bibfnamefont {J.~K.}\ \bibnamefont
  {Hale}},\ }\href@noop {} {\bibfield  {journal} {\bibinfo  {journal} {Journal
  of Dynamics and Differential Equations}\ }\textbf {\bibinfo {volume} {9}},\
  \bibinfo {pages} {1} (\bibinfo {year} {1997})}\BibitemShut {NoStop}%
\bibitem [{\citenamefont {Box}\ \emph {et~al.}(2008)\citenamefont {Box},
  \citenamefont {Jenkins},\ and\ \citenamefont
  {Reinsel}}]{Box+Jenkins+Reinsel:2008}%
  \BibitemOpen
  \bibfield  {author} {\bibinfo {author} {\bibfnamefont {G.~E.~P.}\
  \bibnamefont {Box}}, \bibinfo {author} {\bibfnamefont {G.~M.}\ \bibnamefont
  {Jenkins}}, \ and\ \bibinfo {author} {\bibfnamefont {G.~C.}\ \bibnamefont
  {Reinsel}},\ }\href@noop {} {\emph {\bibinfo {title} {Time Series Analysis:
  Forecast and Control}}},\ \bibinfo {edition} {4th}\ ed.,\ Wiley series in
  probability and statistics\ (\bibinfo  {publisher} {John Wiley \& Sons},\
  \bibinfo {year} {2008})\BibitemShut {NoStop}%
\bibitem [{\citenamefont {Pindyck}\ and\ \citenamefont
  {Rubinfeld}(1981)}]{Pindyck+Rubinfeld:1981}%
  \BibitemOpen
  \bibfield  {author} {\bibinfo {author} {\bibfnamefont {R.}~\bibnamefont
  {Pindyck}}\ and\ \bibinfo {author} {\bibfnamefont {D.}~\bibnamefont
  {Rubinfeld}},\ }\href@noop {} {\emph {\bibinfo {title} {Econometric Models
  and Economic Forecasts}}},\ Vol.~\bibinfo {volume} {2}\ (\bibinfo
  {publisher} {{McGraw-Hill}},\ \bibinfo {address} {New York},\ \bibinfo {year}
  {1981})\BibitemShut {NoStop}%
\bibitem [{\citenamefont {Haykin}(1999)}]{Haykin:1999}%
  \BibitemOpen
  \bibfield  {author} {\bibinfo {author} {\bibfnamefont {S.}~\bibnamefont
  {Haykin}},\ }\href@noop {} {\emph {\bibinfo {title} {Neural Networks: A
  Comprehensive Foundation}}},\ \bibinfo {edition} {2nd}\ ed.\ (\bibinfo
  {publisher} {Upper Saddle River (NJ): Prentice Hall},\ \bibinfo {year}
  {1999})\BibitemShut {NoStop}%
\bibitem [{\citenamefont {Rasmussen}\ and\ \citenamefont
  {Williams}(2006)}]{Rasmussen:2006}%
  \BibitemOpen
  \bibfield  {author} {\bibinfo {author} {\bibfnamefont {C.~E.}\ \bibnamefont
  {Rasmussen}}\ and\ \bibinfo {author} {\bibfnamefont {C.~K.~I.}\ \bibnamefont
  {Williams}},\ }\href@noop {} {\emph {\bibinfo {title} {Gaussian Processes for
  Machine Learning}}}\ (\bibinfo  {publisher} {The MIT Press},\ \bibinfo {year}
  {2006})\BibitemShut {NoStop}%
\bibitem [{\citenamefont {Qui{\~n}onero-Candela}\ and\ \citenamefont
  {Rasmussen}(2005)}]{Quinonero-Candela+Rasmussen:2005}%
  \BibitemOpen
  \bibfield  {author} {\bibinfo {author} {\bibfnamefont {J.}~\bibnamefont
  {Qui{\~n}onero-Candela}}\ and\ \bibinfo {author} {\bibfnamefont {C.~E.}\
  \bibnamefont {Rasmussen}},\ }\href@noop {} {\bibfield  {journal} {\bibinfo
  {journal} {Journal of Machine Learning Research}\ }\textbf {\bibinfo {volume}
  {6}},\ \bibinfo {pages} {1939} (\bibinfo {year} {2005})}\BibitemShut
  {NoStop}%
\bibitem [{\citenamefont {Lindgren}\ \emph {et~al.}(2011)\citenamefont
  {Lindgren}, \citenamefont {Rue},\ and\ \citenamefont
  {Lindstr\"om}}]{Lindgren+Rue+Linstrom:2011}%
  \BibitemOpen
  \bibfield  {author} {\bibinfo {author} {\bibfnamefont {F.}~\bibnamefont
  {Lindgren}}, \bibinfo {author} {\bibfnamefont {H.}~\bibnamefont {Rue}}, \
  and\ \bibinfo {author} {\bibfnamefont {J.}~\bibnamefont {Lindstr\"om}},\
  }\href@noop {} {\bibfield  {journal} {\bibinfo  {journal} {JRSS B}\ }\textbf
  {\bibinfo {volume} {73}},\ \bibinfo {pages} {423} (\bibinfo {year}
  {2011})}\BibitemShut {NoStop}%
\bibitem [{\citenamefont {S\"arkk\"a}\ and\ \citenamefont
  {Hartikainen}(2012)}]{Sarkka:AISTATS:2012}%
  \BibitemOpen
  \bibfield  {author} {\bibinfo {author} {\bibfnamefont {S.}~\bibnamefont
  {S\"arkk\"a}}\ and\ \bibinfo {author} {\bibfnamefont {J.}~\bibnamefont
  {Hartikainen}},\ }in\ \href@noop {} {\emph {\bibinfo {booktitle} {JMLR
  Workshop and Conference Proceedings Volume 22: AISTATS 2012}}}\ (\bibinfo
  {year} {2012})\ pp.\ \bibinfo {pages} {993--1001}\BibitemShut {NoStop}%
\bibitem [{\citenamefont {Da~Prato}\ and\ \citenamefont
  {Zabczyk}(1992)}]{DaPrato:1992}%
  \BibitemOpen
  \bibfield  {author} {\bibinfo {author} {\bibfnamefont {G.}~\bibnamefont
  {Da~Prato}}\ and\ \bibinfo {author} {\bibfnamefont {J.}~\bibnamefont
  {Zabczyk}},\ }\href@noop {} {\emph {\bibinfo {title} {Stochastic Equations in
  Infinite Dimensions}}},\ \bibinfo {series} {Encyclopedia of Mathematics and
  its Applications}, Vol.~\bibinfo {volume} {45}\ (\bibinfo  {publisher}
  {Cambridge University Press},\ \bibinfo {year} {1992})\BibitemShut {NoStop}%
\bibitem [{\citenamefont {S{\"a}rkk{\"a}}(2013)}]{Sarkka:2013}%
  \BibitemOpen
  \bibfield  {author} {\bibinfo {author} {\bibfnamefont {S.}~\bibnamefont
  {S{\"a}rkk{\"a}}},\ }\href@noop {} {\emph {\bibinfo {title} {Bayesian
  Filtering and Smoothing}}},\ \bibinfo {series} {Institute of Mathematical
  Statistics Textbooks}, Vol.~\bibinfo {volume} {3}\ (\bibinfo  {publisher}
  {Cambridge University Press},\ \bibinfo {year} {2013})\BibitemShut {NoStop}%
\bibitem [{\citenamefont {Chow}(2007)}]{Chow:2007}%
  \BibitemOpen
  \bibfield  {author} {\bibinfo {author} {\bibfnamefont {P.}~\bibnamefont
  {Chow}},\ }\href@noop {} {\emph {\bibinfo {title} {Stochastic Partial
  Differential Equations}}},\ \bibinfo {series} {Chapman \& Hall/CRC applied
  mathematics and nonlinear science series}, Vol.~\bibinfo {volume} {11}\
  (\bibinfo  {publisher} {Chapman \& Hall/CRC Press},\ \bibinfo {year}
  {2007})\BibitemShut {NoStop}%
\bibitem [{\citenamefont {Tzafestas}(1978)}]{Tzafestas:1978}%
  \BibitemOpen
  \bibfield  {author} {\bibinfo {author} {\bibfnamefont {S.~G.}\ \bibnamefont
  {Tzafestas}},\ }in\ \href@noop {} {\emph {\bibinfo {booktitle} {Distributed
  Parameter Systems: Identification, Estimation and Control}}},\ \bibinfo
  {editor} {edited by\ \bibinfo {editor} {\bibfnamefont {W.~H.}\ \bibnamefont
  {Ray}}\ and\ \bibinfo {editor} {\bibfnamefont {D.~G.}\ \bibnamefont
  {Lainiotis}}}\ (\bibinfo  {publisher} {Marcel Dekker, Inc.},\ \bibinfo
  {address} {New York},\ \bibinfo {year} {1978})\BibitemShut {NoStop}%
\bibitem [{\citenamefont {Omatu}\ and\ \citenamefont
  {Seinfeld}(1989)}]{Omatu+Seinfeld:1989}%
  \BibitemOpen
  \bibfield  {author} {\bibinfo {author} {\bibfnamefont {S.}~\bibnamefont
  {Omatu}}\ and\ \bibinfo {author} {\bibfnamefont {J.~H.}\ \bibnamefont
  {Seinfeld}},\ }\href@noop {} {\emph {\bibinfo {title} {Distributed Parameter
  Systems: Theory and Applications}}}\ (\bibinfo  {publisher} {Clarendon Press
  / Ohmsha},\ \bibinfo {year} {1989})\BibitemShut {NoStop}%
\bibitem [{\citenamefont {Cressie}\ and\ \citenamefont
  {Wikle}(2002)}]{Cressie+Wikle:2002}%
  \BibitemOpen
  \bibfield  {author} {\bibinfo {author} {\bibfnamefont {N.}~\bibnamefont
  {Cressie}}\ and\ \bibinfo {author} {\bibfnamefont {C.~K.}\ \bibnamefont
  {Wikle}},\ }in\ \href@noop {} {\emph {\bibinfo {booktitle} {Encyclopedia of
  Environmetrics}}},\ Vol.~\bibinfo {volume} {4},\ \bibinfo {editor} {edited
  by\ \bibinfo {editor} {\bibfnamefont {A.~H.}\ \bibnamefont {El-Shaarawi}}\
  and\ \bibinfo {editor} {\bibfnamefont {W.~W.}\ \bibnamefont {Piegorsch}}}\
  (\bibinfo  {publisher} {John Wiley \& Sons, Ltd, Chichester},\ \bibinfo
  {year} {2002})\ pp.\ \bibinfo {pages} {2045--2049}\BibitemShut {NoStop}%
\bibitem [{\citenamefont {Grewal}\ and\ \citenamefont
  {Andrews}(2001)}]{Grewal+Andrews:2001}%
  \BibitemOpen
  \bibfield  {author} {\bibinfo {author} {\bibfnamefont {M.~S.}\ \bibnamefont
  {Grewal}}\ and\ \bibinfo {author} {\bibfnamefont {A.~P.}\ \bibnamefont
  {Andrews}},\ }\href@noop {} {\emph {\bibinfo {title} {{K}alman Filtering:
  Theory and Practice Using {MATLAB}}}},\ \bibinfo {edition} {2nd}\ ed.\
  (\bibinfo  {publisher} {Wiley-Intersciece},\ \bibinfo {year}
  {2001})\BibitemShut {NoStop}%
\bibitem [{\citenamefont {Singer}(2011)}]{Singer:2011}%
  \BibitemOpen
  \bibfield  {author} {\bibinfo {author} {\bibfnamefont {H.}~\bibnamefont
  {Singer}},\ }\href@noop {} {\bibfield  {journal} {\bibinfo  {journal} {{AStA}
  Advances in Statistical Analysis}\ }\textbf {\bibinfo {volume} {95}},\
  \bibinfo {pages} {375} (\bibinfo {year} {2011})}\BibitemShut {NoStop}%
\bibitem [{Note1()}]{Note1}%
  \BibitemOpen
  \bibinfo {note} {The dataset is available for download from U.S. National
  Climatic Data Center: \protect \url
  {http://www7.ncdc.noaa.gov/CDO/cdoselect.cmd} (accessed December 12,
  2011).}\BibitemShut {Stop}%
\end{thebibliography}%

\end{document}